\documentclass[runningheads]{llncs}
\def\BibTeX{{\rm B\kern-.05em{\sc i\kern-.025em b}\kern-.08em
   T\kern-.1667em\lower.7ex\hbox{E}\kern-.125emX}}
\usepackage{graphicx}
\usepackage{mathtools}
\usepackage[utf8]{inputenc}
\usepackage{lipsum} 
\usepackage{booktabs}
\usepackage[ruled]{algorithm2e}
\usepackage{url}
\usepackage{colortbl}
\usepackage{multicol}
\usepackage{color}
\usepackage{multirow}
\usepackage{enumitem}
\usepackage{array}
\usepackage{siunitx}
\usepackage[labelfont={bf}]{caption}
\usepackage{balance}
\usepackage{graphicx}
\usepackage{array} 
\usepackage{todonotes}
\usepackage{subfig}
\usepackage[nolist,nohyperlinks]{acronym}
\usepackage{footnote}
\usepackage[perpage]{footmisc}
\usepackage[autostyle]{csquotes}
\makesavenoteenv{tabular}

\begin{document}
\title{Transformer based Generative Adversarial Network for Liver Segmentation}

\author{Ugur Demir*  \and
Zheyuan Zhang* \and
Bin Wang \and
Matthew Antalek \and
Elif Keles  \and
Debesh Jha \and
Amir Borhani \and
Daniela Ladner\and
Ulas Bagci}
\institute{Northwestern University, IL 60201, USA \\
\thanks{Those authors contribute equally to this paper.}
\email{ulas.bagci@northwestern.edu}}
\authorrunning{Demir, Zhang et al.}

\maketitle             
\begin{abstract}
Automated liver segmentation from radiology scans (CT, MRI) can improve surgery and therapy planning and follow-up assessment in addition to conventional use for diagnosis and prognosis. Although convolutional neural networks (CNNs) have became the standard image segmentation tasks, more recently this has started to change towards Transformers based architectures because Transformers are  taking advantage of capturing long range dependence modeling capability in signals, so called attention mechanism.  In this study, we propose a new segmentation approach using a hybrid approach combining the Transformer(s) with the Generative Adversarial Network (GAN) approach. The premise behind this choice is that the self-attention mechanism of the Transformers allows the network to aggregate the high dimensional feature and provide global information modeling. This mechanism provides better segmentation performance compared with traditional methods. Furthermore, we encode this generator into the GAN based architecture so that the discriminator network in the GAN can classify the credibility of the generated segmentation masks compared with the real masks coming from human (expert) annotations. This allows us to extract the high dimensional topology information in the mask for biomedical image segmentation and provide more reliable segmentation results. Our model achieved a high dice coefficient of 0.9433, recall of 0.9515, and precision of 0.9376 and outperformed other Transformer based approaches. The implementation details of the proposed architecture can be found at \url{https://github.com/UgurDemir/tranformer_liver_segmentation}. 

\end{abstract}

\section{Introduction}
Liver cancer is among the leading causes of cancer-related deaths, accounting for 8.3\% of cancer mortality~\cite{sung2021global}. The high variability in shape, size, appearance, and local orientations makes liver (and liver diseases such as tumors, fibrosis) challenging to analyze from radiology scans for which the image segmentation is often necessary~\cite{chlebus2018automatic}. An accurate organ and lesion segmentation  could facilitate reliable diagnosis and therapy planning including prognosis ~\cite{cornelis2017precision}. 

As a solution to biomedical image segmentation, the literature is vast and rich. The self-attention mechanism is nowadays widely used in the biomedical image segmentation field where long-range dependencies and context dependent features are essential. By capturing such information, transformer based segmentation architectures (for example, SwinUNet~\cite{cao2021swinunet}) have achieved promising performance on many vision tasks including biomedical image segmentation~\cite{huang2021missformer,vaswani2017attention}.

In parallel to the all advances in Transformers, generative methods have achieved remarkable progresses in almost all fields of computer vision too~\cite{maria}. For example, Generative Adversarial Networks (GAN)~\cite{goodfellow2014generative} is a widely used tool for generating one target image from one source image. GAN has been  applied to the image segmentation framework to distinguish the credibility of the generated masks like previous studies \cite{luc:hal-01398049,najiPan2019}. The high dimensional topology information is an important feature for pixel levell classification, thus segmentation. For example, the segmented mask should recognize the object location, orientation, and scale  prior to delineation procedure, but most current segmentation engines are likely to provide false positives outside the target region or conversely false negatives within the target region due to an inappropriate recognition of the target regions. By introducing the discriminator architecture (as a part of GAN) to distinguish whether the segmentation mask is high quality or not, we could proactively screen poor predictions from the segmentation model. Furthermore, this strategy can also allow us to take advantage of many unpaired segmentation masks which can be easily acquired or even simulated in the segmentation targets. To this end, in this paper, we propose a Transformer based GAN architecture as well as a Transformer based CycleGAN architecture for automatic liver segmentation, a very improtant clinical precursor for liver diseases. By combining two strong algorithms, we aim to achieve both good recognition (localization) of the target region and high quality delineations. 


\section{Proposed method}
\begin{figure} [!t]
    \centering
    \includegraphics[width= 0.9\linewidth]{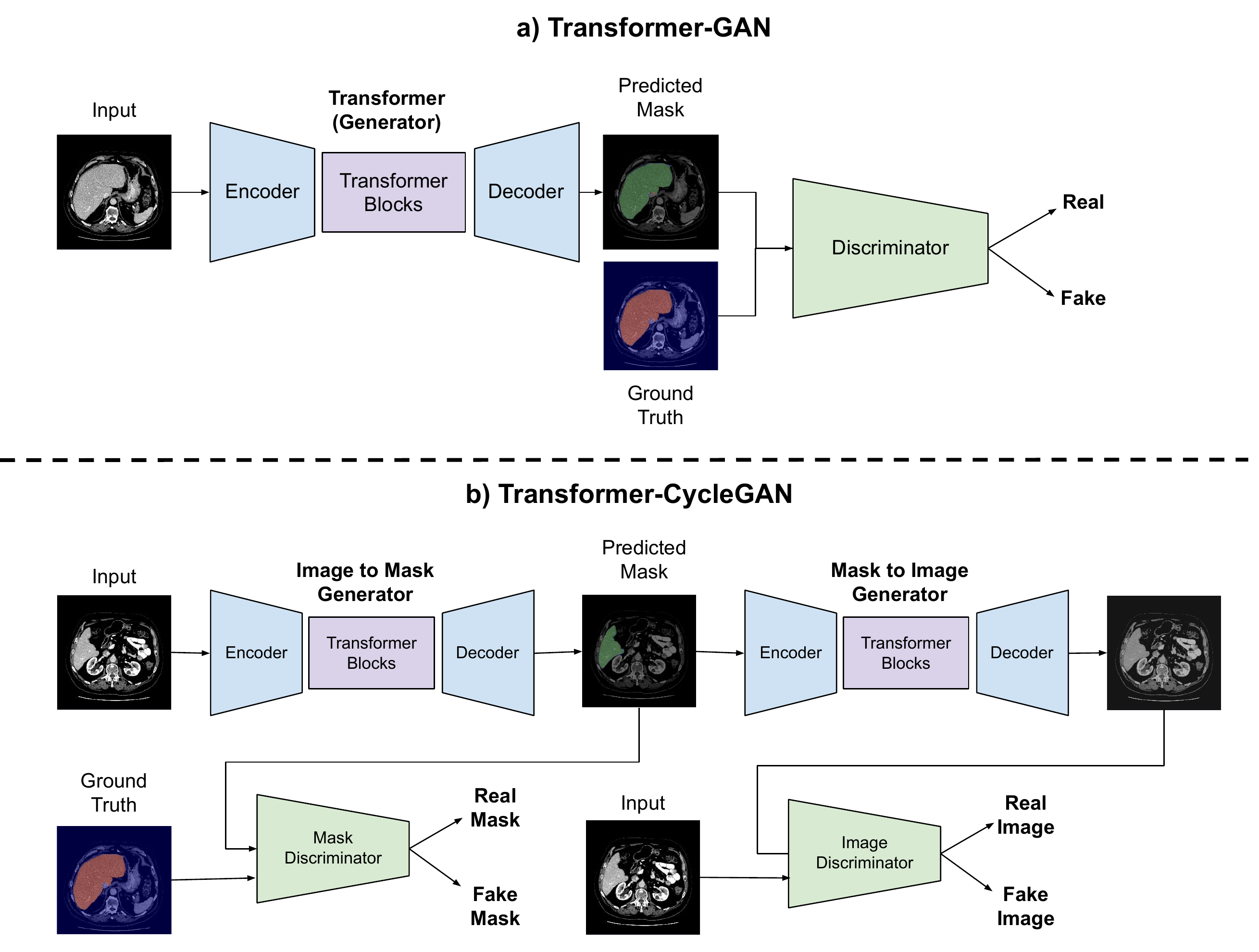}
    \caption{Block diagram of the Transformer GAN. (a) Vanilla GAN and (b) Cycle-GAN with Transformer generator architectures. }
    
    \label{fig:arch}
\end{figure}

We first investigated the transformer architecture to solve the liver segmentation problem from radiology scans, CT in particular due to its widespread use and being the first choice in most liver disease quantification. The self-attention mechanism of the Transformers has been demonstrated to be very effective approach when finding long range dependencies as stated before. This can be quite beneficial for the liver segmentation problem especially because the object of interest (liver) is large and pixels constituting the same object are far from each other. We also utilized an adversarial training approach to boost the segmentation model performance. For this, we have devised a conditional image generator in a vanilla-GAN that learns a mapping between the CT slices and the segmentation maps (i.e., surrogate of the truths or reference standard). The adversarial training forces the generator model to predict more realistic segmentation outcomes. In addition to vanilla-GAN, we have also utilized the CycleGAN~\cite{zhu2017unpaired},~\cite{ristea2021cytran} approach to investigate the effect of cycle consistency constraint on the segmentation task. Figure~\ref{fig:arch} demonstrates the general overview of the proposed method.


\subsection{Transformer based GAN}
Like other GAN architectures~\cite{liu2019cross}, Transformer based GAN architecture is composed of two related sub-architectures: the generator and the discriminator. The generator part could generate the segmentation mask from the raw image (i.e., segmentation task itself), while the discriminator tries to distinguish predictions from the human annotated ground truth. GAN provides a better way to distinguish the high-dimensional morphology information. The discriminator can provide the similarity between the predicted masks and the ground truth (i.e., surrogate truth) masks. Vanilla GAN considers the whole segmentation to decide whether it is fake or not. 

\subsection{Transformer based CycleGAN}
One alternative extension to the standard GAN approach is to use transformer based segmentation model within the CycleGAN setup. Unlike a standard GAN, CycleGAN consists of two generators and two discriminator networks. While the first generator accepts the raw images as input and predicts the segmentation masks, the second generator takes the predicted segmentation maps as input and maps them back to the input image. The first discriminator classifies the segmentation masks as either real or fake, and the second discriminator distinguishes the real and the reconstructed image. Figure~\ref{fig:arch} illustrates this procedure  with liver segmentation from CT scans.  

%

To embed transformers within the CycleGAN,  we utilized the encoder-decoder style convolutional transformer model~\cite{ristea2021cytran}. The premise behind this idea was that the encoder module takes the input image and decreases the spatial dimensions while extracting features with convolution layers. This allowed processing of large-scale images. The core transformer module consisted of several stacked linear layers and self-attention blocks. The decoder part increased the spatial dimension of the intermediate features and makes the final prediction. For the discriminator network, we tried three convolutional architectures. The vanilla-GAN discriminator evaluates the input image as a whole. Alternatively, we have adopted PatchGAN discriminator architecture~\cite{pix2pix2017} to focus on small mask patches to decide the realness of each region. It splits the input masks into NxN regions and asses their quality individually. When we set the patch size to a pixel, PatchGAN can be considered as pixel level discriminator. W have observed that the pixel level discriminator tends to surpass other architecture for segmentation. Figure~\ref{fig:arch} demonstrates the network overview. In all of the experiments, the segmentation model uses the same convolutional transformer and pixel level discriminator architectures.


\section{Experimental setup}
We have used Liver Tumor Segmentation Challenge (LiTS)\cite{bilic2019liver} dataset. LiTS consists of 131 CT scans. This dataset is publicly available under segmentation challenge website and approved IRB by the challenge organizers. More information about the dataset and challenge can be found here \footnote{\url{https://competitions.codalab.org/competitions/17094#learn_the_details}}. 

All our models were trained on NVIDIA RTX A6000 GPU after implemented using the PyTorch~\cite{paszke2019pytorch} framework. We have used 95 samples for training and 36 samples for testing. All models are trained on the same hyperparameters configuration with a learning rate of $2e^{-4}$, and Adam optimizer with beta1 being 0.5 and beta2 being 0.999. All of the discriminators use the pixel level discriminator in both GAN and CycleGAN experiments. We have used recall, precision, and dice coefficient for quantitative evaluations of the segmentation. Further, segmentation results were qualitatively evaluated by the participating physicians. Our algorithms are available for public use.

\begin{figure} [!t]
    \centering
    \includegraphics[width= 0.95\linewidth]{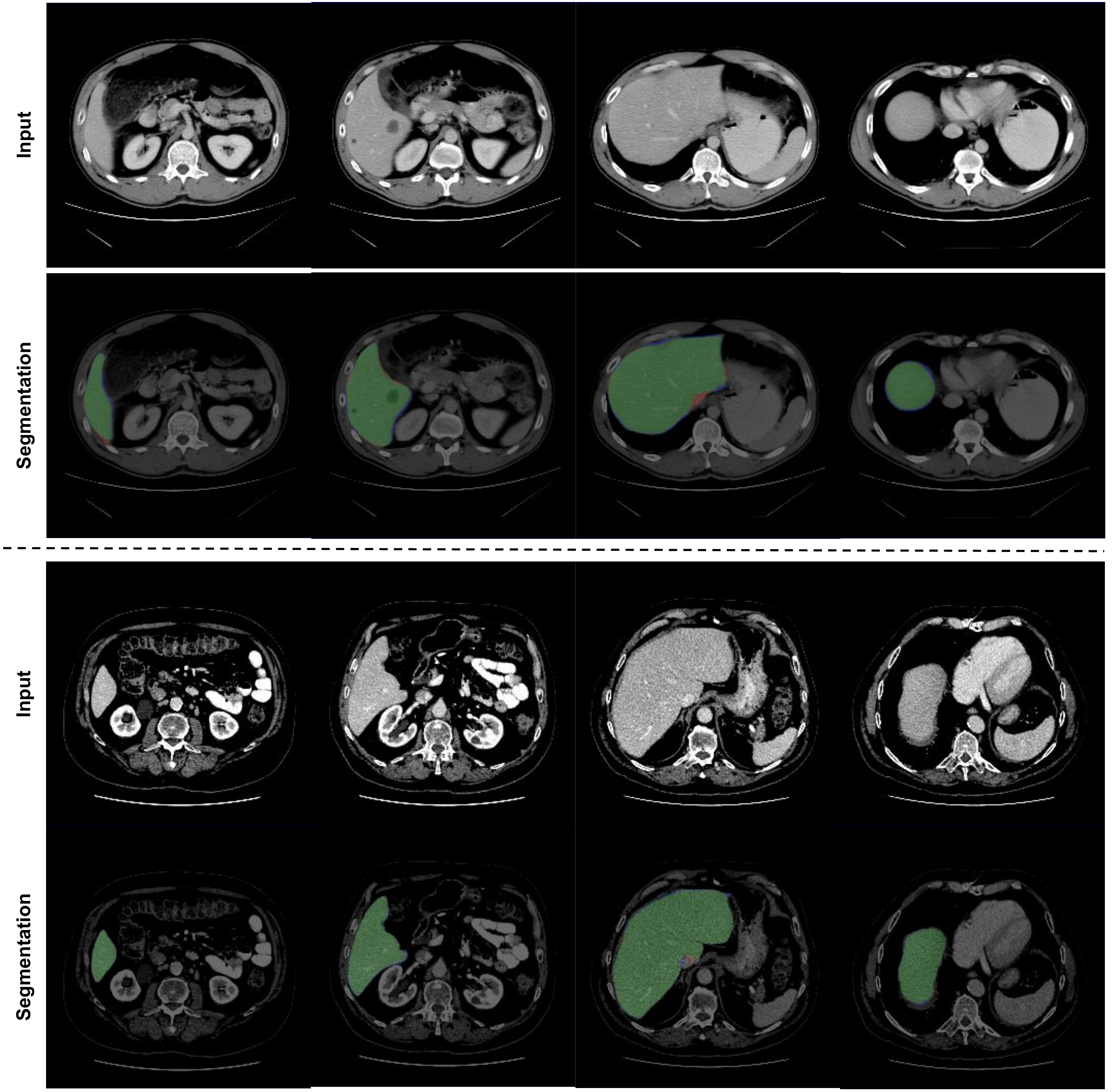}
    \caption{Transformer based GAN liver segmentation results. {\color{green}Green}: True positive, {\color{red}Red}: False Positive, {\color{blue}Blue}: False Negative.
 }
    \label{fig:samples}
\end{figure}

\begin{table}[!t]
\centering
\caption{Performance of Transformer based methods on the LITS dataset.~\cite{bilic2019liver}}
\label{tab1}
\begin{tabular}{l|c|c|c}
\toprule
\textbf{Method} & \textbf{Dice coefficient} & \textbf{Precision} & \textbf{Recall} \\
\hline

Transformer~\cite{Wu_2021_ICCV,ristea2021cytran} & 0.9432 & 0.9464 & 0.9425\\ 
Transformer - CycleGAN (ours) & 0.9359 & \textbf{0.9539} & 0.9205 \\
Transformer - GAN (ours)& \textbf{0.9433} & 0.9376 & \textbf{0.9515} \\ 
\bottomrule
\end{tabular}
\end{table}

\section{Results}
We presented the evaluation results in Table~\ref{tab1}. Our best performing method was Transformer based GAN architecture, achieved a highest dice coefficient of 0.9433 and recall rate of 0.9515. Similarly, our transformer based CycleGAN architecture has the highest precision, 0.9539. With Transformer based GAN, we achieved 0.9\% improvement in recall and 0.01\% improvement in dice coefficient with respect to the vanilla Transformers. It is to be noted that we have used also post-processing technique which boosts the performance for "all" the baselines to avoid biases one from each other.

Figure \ref{fig:samples} shows our qualitative results for the liver segmentation. We have examined all the liver segmentation results one-by-one and no failure were identified by the participating physicians. Hence, visual results agreed with the quantitative results as described in Table 1.

\section{Conclusion}
In this study, we explored the use of transformer-based GAN architectures for medical image segmentation. Specifically, we used a self-attention mechanism and designed a discriminator for classifying the credibility of generated segmentation masks. Our experimental result showed that the proposed new segmentation architectures could provide accurate and reliable segmentation performance as compared to the baseline Transfomers. Although we have shown our results in an important clinical problem for liver diseases where image-based quantification is vital, the proposed hybrid architecture (i.e., combination of GAN and Transformers) can potentially be applied to various medical image segmentation tasks beyond liver CTs as the algorithms are generic, reproducible, and carries similarities with the other segmentation tasks in biomedical imaging field. We anticipate that our architecture can also be applied to medical scans within the semi-supervised learning, planned as a future work.

\section*{Acknowledgement}
This study is partially supported by NIH NCI grants R01-CA246704 and R01-R01-CA240639.

\bibliographystyle{splncs04}
\bibliography{mybibliography}

\begin{thebibliography}{10}
\providecommand{\url}[1]{\texttt{#1}}
\providecommand{\urlprefix}{URL }
\providecommand{\doi}[1]{https://doi.org/#1}

\bibitem{bilic2019liver}
Bilic, P., Christ, P.F., Vorontsov, E., Chlebus, G., Chen, H., Dou, Q., Fu,
  C.W., Han, X., Heng, P.A., Hesser, J., et~al.: The liver tumor segmentation
  benchmark (lits). arXiv preprint arXiv:1901.04056  (2019)

\bibitem{cao2021swinunet}
Cao, H., Wang, Y., Chen, J., Jiang, D., Zhang, X., Tian, Q., Wang, M.:
  Swin-unet: Unet-like pure transformer for medical image segmentation. arXiv
  preprint arXiv:2105.05537  (2021)

\bibitem{chlebus2018automatic}
Chlebus, G., Schenk, A., Moltz, J.H., van Ginneken, B., Hahn, H.K., Meine, H.:
  Automatic liver tumor segmentation in ct with fully convolutional neural
  networks and object-based postprocessing. Scientific reports  \textbf{8}(1),
  ~1--7 (2018)

\bibitem{maria}
Chuquicusma, M.J., Hussein, S., Burt, J., Bagci, U.: How to fool radiologists
  with generative adversarial networks? a visual turing test for lung cancer
  diagnosis. In: 2018 IEEE 15th international symposium on biomedical imaging
  (ISBI 2018). pp. 240--244. IEEE (2018)

\bibitem{cornelis2017precision}
Cornelis, F., Martin, M., Saut, O., Buy, X., Kind, M., Palussiere, J., Colin,
  T.: Precision of manual two-dimensional segmentations of lung and liver
  metastases and its impact on tumour response assessment using recist 1.1.
  European radiology experimental  \textbf{1}(1), ~1--7 (2017)

\bibitem{goodfellow2014generative}
Goodfellow, I., Pouget-Abadie, J., Mirza, M., Xu, B., Warde-Farley, D., Ozair,
  S., Courville, A., Bengio, Y.: Generative adversarial nets. Advances in
  neural information processing systems  \textbf{27} (2014)

\bibitem{huang2021missformer}
Huang, X., Deng, Z., Li, D., Yuan, X.: Missformer: An effective medical image
  segmentation transformer. arXiv preprint arXiv:2109.07162  (2021)

\bibitem{pix2pix2017}
Isola, P., Zhu, J.Y., Zhou, T., Efros, A.A.: Image-to-image translation with
  conditional adversarial networks. CVPR  (2017)

\bibitem{najiPan2019}
Khosravan, N., Mortazi, A., Wallace, M., Bagci, U.: {PAN: Projective
  Adversarial Network for Medical Image Segmentation}. In: {Medical Image
  Computing and Computer Assisted Intervention – MICCAI 2019 - 22nd
  International Conference, Proceedings} (2019)

\bibitem{liu2019cross}
Liu, Y., Khosravan, N., Liu, Y., Stember, J., Shoag, J., Bagci, U.,
  Jambawalikar, S.: Cross-modality knowledge transfer for prostate segmentation
  from ct scans. In: Domain adaptation and representation transfer and medical
  image learning with less labels and imperfect data, pp. 63--71. Springer
  (2019)

\bibitem{luc:hal-01398049}
Luc, P., Couprie, C., Chintala, S., Verbeek, J.: {Semantic Segmentation using
  Adversarial Networks}. In: {NIPS Workshop on Adversarial Training}.
  Barcelona, Spain (Dec 2016), \url{https://hal.inria.fr/hal-01398049}

\bibitem{paszke2019pytorch}
Paszke, A., Gross, S., Massa, F., Lerer, A., Bradbury, J., Chanan, G., Killeen,
  T., Lin, Z., Gimelshein, N., Antiga, L., et~al.: Pytorch: An imperative
  style, high-performance deep learning library. Advances in neural information
  processing systems  \textbf{32} (2019)

\bibitem{ristea2021cytran}
Ristea, N.C., Miron, A.I., Savencu, O., Georgescu, M.I., Verga, N., Khan, F.S.,
  Ionescu, R.T.: Cytran: Cycle-consistent transformers for non-contrast to
  contrast ct translation. arXiv preprint arXiv:2110.06400  (2021)

\bibitem{sung2021global}
Sung, H., Ferlay, J., Siegel, R.L., Laversanne, M., Soerjomataram, I., Jemal,
  A., Bray, F.: Global cancer statistics 2020: Globocan estimates of incidence
  and mortality worldwide for 36 cancers in 185 countries. CA: a cancer journal
  for clinicians  \textbf{71}(3),  209--249 (2021)

\bibitem{vaswani2017attention}
Vaswani, A., Shazeer, N., Parmar, N., Uszkoreit, J., Jones, L., Gomez, A.N.,
  Kaiser, {\L}., Polosukhin, I.: Attention is all you need. Advances in neural
  information processing systems  \textbf{30} (2017)

\bibitem{Wu_2021_ICCV}
Wu, H., Xiao, B., Codella, N., Liu, M., Dai, X., Yuan, L., Zhang, L.: Cvt:
  Introducing convolutions to vision transformers. In: Proceedings of the
  IEEE/CVF International Conference on Computer Vision (ICCV). pp. 22--31
  (October 2021)

\bibitem{zhu2017unpaired}
Zhu, J.Y., Park, T., Isola, P., Efros, A.A.: Unpaired image-to-image
  translation using cycle-consistent adversarial networks. In: Proceedings of
  the IEEE international conference on computer vision. pp. 2223--2232 (2017)

\end{thebibliography}

\end{document}